\documentclass[12pt,eadjointtfnpsf]{article}

\usepackage{amsmath,amsbsy,amssymb,latexsym,epsfig}

\newcommand{\BE}{\begin{equation}}
\newcommand{\EE}{\end{equation}}
\newcommand{\BEA}{\begin{eqnarray}}
\newcommand{\EEA}{\end{eqnarray}}

\def\12{\frac{1}{2}}
\def\bea{\begin{eqnarray}}
\def\eea{\end{eqnarray}}
\def\ba{\begin{array}}
\def\ea{\end{array}}

\def\one-loop{\mbox{\scriptsize one-loop}}

\def\G{\Gamma}

\catcode`\@=11

\@addtoreset{equation}{section}
\def\theequation{\arabic{section}.\arabic{equation}}
\def\@normalsize{\@setsize\normalsize{15pt}\xiipt\@xiipt
\abovedisplayskip 14pt plus3pt minus3pt%
\belowdisplayskip \abovedisplayskip
\abovedisplayshortskip \z@ plus3pt%
\belowdisplayshortskip 7pt plus3.5pt minus0pt}

\def\small{\@setsize\small{13.6pt}\xipt\@xipt
\abovedisplayskip 13pt plus3pt minus3pt%
\belowdisplayskip \abovedisplayskip
\abovedisplayshortskip \z@ plus3pt%
\belowdisplayshortskip 7pt plus3.5pt minus0pt
\def\@listi{\parsep 4.5pt plus 2pt minus 1pt
\itemsep \parsep
\topsep 9pt plus 3pt minus 3pt}}

\def\underline#1{\relax\ifmmode\@@underline#1\else
$\@@underline{\hbox{#1}}$\relax\fi}
\@twosidetrue

\relax

\catcode`@=12

\catcode`\@=11

\def\section{\@startsection{section}{1}{\z@}{3.5ex plus 1ex minus
.2ex}{2.3ex plus .2ex}{\large\bf}}

\def\thesection{\Roman{section}.}

\def\appendix{\setcounter{section}{0}
\def\thesection{Appendix }

\def\theequation{\Alph{section}.\arabic{equation}}}

\catcode`\@=12

\relax

\def\figcap{\section*{Figure Captions\markboth
{FIGURECAPTIONS}{FIGURECAPTIONS}}\list
{Fig. \arabic{enumi}:\hfill}{\settowidth\labelwidth{Fig. 999:}
\leftmargin\labelwidth
\advance\leftmargin\labelsep\usecounter{enumi}}}
 \relax
\def\tablecap{\section*{Table Captions\markboth
{TABLECAPTIONS}{TABLECAPTIONS}}\list
{Table \arabic{enumi}:\hfill}{\settowidth\labelwidth{Table 999:}
\leftmargin\labelwidth
\advance\leftmargin\labelsep\usecounter{enumi}}}
 \relax
\def\reflist{\section*{References\markboth
{REFLIST}{REFLIST}}\list
{[\arabic{enumi}]\hfill}{\settowidth\labelwidth{[999]}
\leftmargin\labelwidth
\advance\leftmargin\labelsep\usecounter{enumi}}}
 \relax

\newskip\humongous \humongous=0pt plus 1000pt minus 1000pt

\newif\ifdtup

\def\beq{\begin{equation}}
\def\eeq{\end{equation}}

\def\beqn{\begin{eqnarray}}
\def\eeqn{\end{eqnarray}}
\relax

\def\G2{{\; \rm GeV/}c2}
\def\G{\; \rm GeV}

\def\dotx{\dotx{\dot\overline{x}}}

\relax

\renewcommand{\thefootnote}{\fnsymbol{footnote}}

\begin{document}
%
%
\begin{titlepage}

\begin{flushright}
\normalsize
September, 2005 \\
OCU-PHYS 233 \\
hep-th/0509146 \\
\end{flushright}

\begin{center}
{\large\bf  
Matrix Orientifolding and \\Models with Four or Eight Supercharges}
\end{center}

\vfill

\begin{center}
{%
H. Itoyama\footnote{e-mail: itoyama@sci.osaka-cu.ac.jp},
\quad \quad
R. Yoshioka\footnote{e-mail: yoshioka@sci.osaka-cu.ac.jp}
}

\end{center}

\vfill

\begin{center}
\it Department of Mathematics and Physics,
Graduate School of Science\\
Osaka City University\\
\medskip

\bigskip

3-3-138, Sugimoto, Sumiyoshi, Osaka, 558-8585, Japan \\

\end{center}

\vfill

\begin{abstract}
The conditions under which matrix orientifolding and supersymmetry
 transformations commute are known to be stringent. 
Here we present the cases possessing four  or eight
 supercharges upon ${\bf Z}_3$  orbifolding followed by
 matrix orientifolding. These cases descend from the matrix models
 with  eight plus eight supercharges.  There are fifty
 in total, which we enumerate.
\end{abstract}

\vfill
\setcounter{footnote}{0}
\renewcommand{\thefootnote}{\arabic{footnote}}

\end{titlepage}
\section{Introduction}
\label{1}

 Continuing attention has been paid to matrix models which 
 are proposed to enable nonperturbative studies of strings
  beyond their perturbative and semiclassical regimes.\cite{BFSS,IKKT,IT2,IT1,DF,KSlv}
 The objects playing a central role are, of course, 
 discretized  string coordinates represented by matrices
  taking values of appropriate Lie algebras.
 Their diagonal entries represent spacetime points,
 while off-diagonal ones mediate interactions between blocks 
 which may be identified as D-objects.
 A few  ideas on the formation of our spacetime
 such as  the one via branched polymers\cite{AIKKT} and the one  via
 generalized monopoles\cite{CIK2} have appeared  and 
 approximation schemes\cite{NS} have been devised.(See \cite{T} for more references.)

Not only the string coordinates  but also algebraic operations in
 the first quantized string theory 
 have natural matrix counterparts even when the size 
of the matrices is kept finite. In particular,
 the matrix counterpart of  twist operation or orientifolding
  is easily obtained as  any $U(2k)$ Lie algebra  valued matrix splits
 into  a direct sum of
   the adjoint representation and  the antisymmetric representation of
 $USp(2k)$ or $SO(2k)$ Lie algebra.
 Selecting one of these two representations  for each of the
 original matrix coordinates is  referred to as 
 matrix orientifolding in this paper.

  Realizing the twist operation of matrices this way has turned out to
 put stringent conditions on the number of supercharges\cite{IT1}:
 the supersymmetry transformations in the Wess-Zumino gauge are
  non-linear and  requiring that they commute with the projectors
  materializing matrix orientifolding yields nontrivial
 algebraic conditions. 
 In the case of $8+8$ supercharges, these conditions are
 successful in selecting the two known cases which
  corresponds to the $USp$ matrix model\cite{IT2,IT1} relevant to type I superstrings
 and  the matrix model\cite{DF,KSlv} of heterotic M theory\cite{HW}.
 In the light of assessing these algebraic conditions further and
 of hoping to find principal matrix configurations  leading to
 ${\cal N}=1$ vacua in four dimensions, 
 it is  interesting to find out how many cases of matrix orientifolding 
  one can construct which possess fewer supercharges.
  To put this question more concrete,  consider
  the matrix analog  of ${\bf C}^{3}/{\bf Z}_{3}$\cite{AIS}, 
 and subsequently operate matrix orientifolding.
  In this paper, we focus upon the problem of
 enumerating all possible such cases with supersymmetries, namely,
  the ones obtained by 
 ${\bf Z}_{3}$ orbifolding  followed by matrix orientifolding.

In the next section, we recall the two cases of matrix orientifolding
 with $8+8$ supercharges.
After introducing ${\bf Z}_{3}$ orbifolding
 acting upon three complex matrix coordinates and its prototypical
 example in section III, we carry out the matrix
 orientifolding of this example in section IV.
  We show that  there are two consistent possibilities with respect to
 supersymmetries and that there are in total five cases:
  the one possesses $4+0$ supersymmetries while the remaining four possess
  $2+2$ supersymmetries.
 The problem to enumerate all cases obtained upon an arbitrary
 ${\bf Z}_{3}$ orbifolding and subsequently  matrix orientifolding
 while keeping some supersymmetries intact 
 is addressed in section V.   ${\bf Z}_{3}$ orbifolding leaves either
  $4+4$ supersymmetries or $8+8$ supersymmetries intact.
  We show that, to each  of the four possibilities  belonging to the former,
  there is one case of consistent matrix orientifolding with $4+0$
  supersymmetries and  four  with $2+2$ supersymmetries.
As for each of the six possibilities belonging to the latter 
($8+8$ supersymmetries), we show that there is also one with $8+0$ 
supersymmetries and four with $4+4$ supersymmetries.  
The total number of such cases is fifty.  
This number is considered to be small in  the light of
 an innumerable number of perturbative superstring vacua.

\section{Matrix Orientifolding with 8+8 Supercharges}
\label{2}

The action of the IIB matrix model is
\beq
S=-\frac{1}{g^2}\text{Tr}(\frac{1}{4}[A_N,A_M][A^N,A^M]+
\frac{1}{2}\bar{\psi}\Gamma^N[A_N,\psi]) .
\eeq
Here $\psi$ is a ten-dimensional Majorana-Weyl spinor, and
$A_I$ and $\psi$ are $N\times N$ Hermitian matrices.
The action has  dynamical supersymmetry
\begin{align}
&\delta^{(1)}\psi=\frac{i}{2}[A_N,A_M]\Gamma^{NM}\epsilon, \\
&\delta^{(1)}A_N=i\bar{\epsilon}\Gamma^N\psi ,
\end{align}
  and  kinematical supersymmetry
\begin{align}
&\delta^{(2)}\psi=\xi, \\
&\delta^{(2)}A_N=0.
\end{align}
As is mentioned in the introduction, any $U(2k)$ Lie algebra valued
 matrix  splits into a direct sum of the two matrices
 which are respectively  the adjoint representation and the antisymmetric
representation of $USp(2k)$ Lie algebra  and this is schematically drawn as
\[U(2k)\ \text{adjoint}\ 
\begin{matrix}
\stackrel{\hat{\rho}_-}{\nearrow}&\text{USp\ adjoint}\\
\stackrel{\hat{\rho}_+}{\searrow}&\text{USp\ antisymmetric}
\end{matrix}\]
\begin{align}
&\text{adj\ }X:\ X^tF+FX=0\label{adj}\\
&\text{asy\ }Y:\ Y^tF-FY=0\label{asy}.
\end{align}
Here $F$ is the matrix counterpart of the twist operation
\begin{equation}
F=\begin{pmatrix}
0&I_k\\-I_k&0\end{pmatrix},
\end{equation}
 and  $\hat{\rho}_{\mp}$  are the projectors
\begin{equation}
\hat{\rho}_{\mp}\bullet=\frac{1}{2}(\bullet\mp F^{-1}\bullet^tF).
\end{equation}

Let
\begin{align}
v_M&\equiv\delta_M^N\hat{\rho}_{b\mp}^{(N)}A_N,\notag\\
\Psi_A&\equiv\delta_{AB}\hat{\rho}_{f\mp}^{(B)}\psi_B,
\label{pro}
\end{align}
where $\hat{\rho}^{(N)}_{b\mp}$ and $\hat{\rho}^{(B)}_{f\mp}$ are
either $\hat{\rho}_-$ or $\hat{\rho}_+$ for each $N$ and for each $B$ respectively.
More explicitly 
\begin{gather}
\hat{\rho}^{(M)}_{b\mp}\equiv\Theta( M\in\mathcal{M}_-)\hat{\rho}_-
+\Theta(M\in\mathcal{M}_+)\hat{\rho}_+, \notag\\
\hat{\rho}^{(A)}_{f\mp}\equiv\Theta(A\in\mathcal{A}_-)\hat{\rho}_-
+\Theta (M\in\mathcal{A}_+)\hat{\rho}_+,
\end{gather}
where
\begin{gather}
\mathcal{M}_-\cup\mathcal{M}_+=\{\{0,1,2,3,4,5,6,7,8,9\}\},
\mspace{20mu}
\mathcal{M}_-\cap\mathcal{M}_+=\emptyset,\\
\mathcal{A}_-\cup\mathcal{A}_+=\{\{1,2,5,6,9,10,13,14,19,20,23,24,27,28,31,32\}\},
\mspace{20mu}
\mathcal{A}_-\cap\mathcal{A}_+=\emptyset.
\end{gather}
By construction, each component of $v_M$ and that of $\Psi_A$
belong either to the adjoint or to the antisymmetric representation of USp(2k).
We impose eq.(\ref{pro}) on $A_N$ and $\psi_B$.
The condition $[\hat{\rho}_{b\mp},\delta^{(1)}]A=0$ gives
\begin{equation}
\sum_A(\bar{\epsilon}\Gamma_M)_A(\hat{\rho}^{(A)}_{f\mp}-\hat{\rho}^{(M)}_{b\mp})\psi_A=0
\label{b-cond}
\end{equation}
with $M$ not summed, while the condition
$[\hat{\rho}_{f\mp},\delta^{(1)}]\psi|_{v_M\to\hat{\rho}_{b\mp}v_M}=0$ gives
\begin{equation}
(1-\hat{\rho}^{(A)}_{f\mp})[\hat{\rho}^{(M)}_{b\mp}A_M,\hat{\rho}^{(N)}_{b\mp}A_N]
(\Gamma^{MN}\epsilon)_A=0.
\label{f-cond}
\end{equation}
The condition $[\hat{\rho}_{b\mp},\delta^{(1)}]A=0$ does not give
us anything new while $[\hat{\rho}_{f\mp},\delta^{(2)}]\psi=0$ gives
\begin{equation}
\xi_A{\bf1}=\xi_A\hat{\rho}^{(A)}_{f\mp}{\bf1}.
\label{k-cond}
\end{equation} 

Eq.(\ref{b-cond}) gives
\begin{equation}
(\bar{\epsilon}\Gamma_{M_{-}})_{A_{+}}=(\bar{\epsilon}\Gamma_{M_{+}})_{A_{-}}=0,
\label{bc}
\end{equation}
while eq.(\ref{f-cond}) gives
\begin{equation}
(\Gamma^{M_-N_+}\epsilon)_{A_-}=0,\mspace{20mu}(\Gamma^{M_-N_-}\epsilon)=
(\Gamma^{M_+N_+}\epsilon)_{A_+}=0,
\label{fc}
\end{equation}
and eq.(\ref{k-cond}) gives
\begin{equation}
\xi_{A_-}=0.
\label{kc}
\end{equation}

Let
\begin{equation}
\epsilon=(\epsilon_0,0,\epsilon_1,0,0,0,0,0,0,\bar{\epsilon}_0,0,\bar{\epsilon}_1,
0,0,0,0)^t.
\end{equation}
The strategy to find solutions to eq.(\ref{bc}), (\ref{fc}) under
 eq.(\ref{b-cond}), (\ref{f-cond}), namely, that of finding two pairs of
 nonintersecting sets $\mathcal{M}_-$ and $\mathcal{M}_+$ and $\mathcal{A}_-$ and
 $\mathcal{A}_+$ are fully described in \cite{IT2} and we will not repeat it here.
The solution is
\begin{gather}
\mathcal{M}_-=\{0,1,2,3,4,7\},
\mspace{20mu}
\mathcal{M}_+=\{5,6,8,9\},\\
\mathcal{A}_-=\{1,2,5,6,19,20,23,24\},
\mspace{20mu}
\mathcal{A}_+=\{9,10,13,14,27,28,31,32\},
\end{gather}
and this leads to  the one\cite{IT2,IT1}\footnote{For further developments of the 
 USp matrix model, see \cite{ITs1,ITs2}. The complete construction of this matrix model
  includes the $n_f=16$ sectors belonging to the (anti-)fundamental representation.
  The use of USp Lie algebra is required by
 the SO($2n_f$) Chan-Paton factor realized by open loop variables.\cite{ITs1,ITs2}}
  of the two known cases 
 of possessing $8+8$ supercharges.
 The corresponding projectors are
\begin{align}
\hat{\rho}_{b\mp}&=\text{diag}(\hat{\rho}_-,\hat{\rho}_-,\hat{\rho}_-,
\hat{\rho}_-,\hat{\rho}_-,\hat{\rho}_+,\hat{\rho}_+,\hat{\rho}_-,
\hat{\rho}_+,\hat{\rho}_+,),\notag\\
\hat{\rho}_{f\mp}&=\hat{\rho}_-I_4\otimes\begin{pmatrix}
I_2&&&\\&0&&\\&&I_2&\\&&&0
\end{pmatrix}
+\hat{\rho}_+I_4\otimes\begin{pmatrix}
0&&&\\&I_2&&\\&&0&\\&&&I_2
\end{pmatrix}.
\end{align}
The other solution\cite{DF,KSlv} with $8+8$ supercharges is
\begin{gather}
\mathcal{M}_-=\{4,7\},
\mspace{20mu}
\mathcal{M}_+=\{0,1,2,3,5,6,8,9\},\\
\mathcal{A}_-=\{1,2,5,6,27,28,31,32\},
\mspace{20mu}
\mathcal{A}_+=\{9,10,13,14,19,20,23,24\}.
\end{gather}
The corresponding projectors are
\begin{align}
\hat{\rho}_{b\mp}&=\text{diag}(\hat{\rho}_+,\hat{\rho}_+,\hat{\rho}_+,
\hat{\rho}_+,\hat{\rho}_-,\hat{\rho}_+,\hat{\rho}_+,\hat{\rho}_-,
\hat{\rho}_+,\hat{\rho}_+,),\notag\\
\hat{\rho}_{f\mp}&=\hat{\rho}_-I_4\otimes\begin{pmatrix}
I_2&&&\\&0&&\\&&0&\\&&&I_2
\end{pmatrix}
+\hat{\rho}_+I_4\otimes
\begin{pmatrix}
0&&&\\&I_2&&\\&&I_2&\\&&&0
\end{pmatrix}.
\end{align}

\section{ ${\bf Z}_3$ Orbifolding}
\label{3}

We now describe  ${\bf Z}_3$ orbifolding of the IIB matrix model.
Let 
   $A_N=(
          A_\mu (\mu=0,\dots,3), 
          B_1=A_4+iA_5,
          B_2=A_6+iA_7,
          B_3=A_8+iA_9 ).$
The complex coordinates $B_i$ are postulated to transform under $\mathbf{Z}_3$ as
\begin{equation}
B_i\to \omega^{a_i}B_i,
\end{equation}
where $a_i$ are integers and $\omega$ is a cubic root of unity.
We introduce the 'tHooft matrices
\begin{equation}
    U=\begin{pmatrix} 1&&\\&\omega&\\&&\omega^2 \end{pmatrix},
    \mspace{20mu}
    V=\begin{pmatrix} 0&0&1\\1&0&0\\0&1&0 \end{pmatrix},
\end{equation}
which satisfy $UV=\omega VU$. 
The $\mathbf{Z}_3$ transformation is given by $M\to UMU^{\dagger}$.
The $\mathbf{Z}_3$ invariant bosonic matrices thus satisfy the conditions;
\begin{equation}
A_\mu=UA_\mu U^{\dagger},
\hspace{5mm}
B_i=\omega^{a_i}UB_iU^{\dagger}.
\end{equation}
In order to find the conditions for the fermionic matrices,
let us note that ten dimensional chirality operator $\Gamma^{10}$
can be thought of as the product of the lower dimensional chirality operators
$\Gamma^{10}=(i\Gamma^0\cdots\Gamma^3)\cdot(i\Gamma^4\Gamma^5)
\cdot(i\Gamma^6\Gamma^7)\cdot(i\Gamma^8\Gamma^9)$ 
and that $\psi$ is  expanded by a set of eigenfunctions $\psi_0\sim\psi_3$,
$\psi_0^c\sim\psi_3^c$ 
\[\psi=\sum_{i=0}^3 \left(\psi_i+(\psi_i)^c\right).\]
The eigenfunctions satisfy the conditions:
\begin{equation}
\psi_i=\omega^{b_i}U\psi_iU^{\dagger},
\end{equation}
where $b_i$ are given by the table
\begin{equation}
\begin{array}{|c||c|c|c|c||c|c|} \hline
\Gamma^{10} & i\Gamma^{0123} & i\Gamma^{45} & i\Gamma^{67} & i\Gamma^{89} & & b_i \\\hline
\multicolumn{1}{|c||}{+} & + & + & + & + & \psi_0 & -(a_1+a_2+a_3)/2 \\\cline{2-7}
\multicolumn{1}{|c||}{} & + & + & - & - & \psi_1 & -(a_1-a_2-a_3)/2 \\\cline{2-7}
\multicolumn{1}{|c||}{} & + & - & + & - & \psi_2 & -(-a_1+a_2-a_3)/2 \\\cline{2-7}
\multicolumn{1}{|c||}{} & + & - & - & + & \psi_3 & -(-a_1-a_2+a_3)/2 \\\cline{2-7}
\multicolumn{1}{|c||}{} & - & - & - & - & (\psi_0)^c & -(-a_1-a_2-a_3)/2 \\\cline{2-7}
\multicolumn{1}{|c||}{} & - & - & + & + & (\psi_1)^c & -(-a_1+a_2+a_3)/2 \\\cline{2-7}
\multicolumn{1}{|c||}{} & - & + & - & + & (\psi_2)^c & -(a_1-a_2+a_3)/2 \\\cline{2-7}
\multicolumn{1}{|c||}{} & - & + & + & - & (\psi_3)^c & -(a_1+a_2-a_3)/2 \\\hline
\end{array}
\label{table}
\end{equation}

The bosonic part of the action is
\begin{equation}
S_b=-\frac{1}{4g^2}\text{Tr}\left([A_\mu,A_\nu]^2+
2\sum_{i=1}^3 [A_\mu,B_i][A^\mu,B_i^{\dagger}]+
\frac{1}{2}\sum_{i,j=1}^3\left([B_i,B_j^{\dagger}][B_i^{\dagger},B_j]+
[B_i,B_j][B_i^{\dagger},B_j^{\dagger}]\right)
\right),
\end{equation}
and the fermionic part is
\begin{equation}
S_f=-\frac{1}{2g^2}\text{Tr}\left(\sum_{i=0}^3\bar{\psi_i}\Gamma^{\mu}[A_{\mu},\psi_i]+
2\sum_{i=1}^3\bar{(\psi_i)^c}\bar{\Gamma}^{(i)}[B_i^{\dagger},\psi_0]+
\sum_{i,j,k=1}^3|\epsilon_{ijk}|\bar{(\psi_i)}^c\Gamma^{(j)}[B_j,\psi_k]+h.c
\right),
\end{equation}
where 
$\Gamma^{(1)}=\frac{1}{2}(\Gamma^4-i\Gamma^5),
\bar{\Gamma}^{(1)}=\frac{1}{2}(\Gamma^4+i\Gamma^5)$ and so on.

A prototypical example is 
\[a_i=2
\mspace{20mu}
\text{for}\ i=1,2,3,\]
and
\[b_0=0 \mspace{10mu}\text{and} \mspace{10mu} b_i=-2=1\mod 3
\mspace{20mu}
\text{for}\ i=1,2,3.\]
Using the 'tHooft matrices, we can represent 
$\mathbf{Z}_3$ invariant matrices $A_{\mu}$, $B_i$, $\psi_0$, and $\psi_i$
as
\begin{equation}
A_\mu=\sum_{a=0}^2A_{\mu}^a\otimes U^a,
B_i=\sum_{a=0}^2B_i^a\otimes(U^aV), 
\psi_0=\sum_{a=0}^2\psi_0^a\otimes U^a,
\psi_i=\sum_{a=0}^2\psi_i^a\otimes(U^aV^{-1}).
\end{equation}

The dynamical supersymmetry is
\begin{align}
\delta^{(1)}\psi_0
&=\frac{i}{2}\left([A_{\mu},A_{\nu}]\Gamma^{\mu\nu}\epsilon_0+[B_i,B_i^{\dagger}]
\epsilon_0\right),\\
\delta^{(1)}\psi_i
&=\frac{i}{2}\left(
|\epsilon_{ijk}|[B_j,B_k]\Gamma^{(j)}\Gamma^{(k)}\epsilon_0
+2[A_{\mu},B_i^{\dagger}]\Gamma^{\mu}\bar{\Gamma}^{(i)}\epsilon_0^c\right),\\
\delta^{(1)}A_{\mu}
&=i\bar{\epsilon_0}\Gamma^{\mu}\psi_0+i\bar{\epsilon_0^c}\Gamma^{\mu}\psi_0^c,\\
\delta^{(1)}B_i
&=2i\bar{\epsilon_0}\bar{\Gamma}^{(i)}\psi_i,
\end{align}
while the kinematical supersymmetry is 
\begin{equation}
\delta^{(2)}\psi_0=\xi_0,
\end{equation} 
and zero otherwise.
This is a model with 4+4 supercharges.

\section{Matrix Orientifolding with Four Supercharges}
\label{4}

Having the discussion of the preceding sections in mind, 
we turn to constructing cases with four supercharges upon 
matrix orientifolding, which descends from
 the case leading to the USp matrix model.
In this section, we restrict our attention to the prototypical
 example of ${\bf Z}_{3}$
 orbifolding discussed in section \ref{3}.
From the condition
$[\hat{\rho}_{f\mp},\delta^{(1)}]\psi_0=0$, 
we obtain
\begin{equation}
(\Gamma^{\mu_-\nu_-}\epsilon_0)_{A_+}=(\Gamma^{\mu_+\nu_+}\epsilon_0)_{A_+}
=0=(\epsilon_0)_{A_+},
(\Gamma^{\mu_-\nu_+}\epsilon_0)_{A_-}=0.
\label{p01}
\end{equation}
Similar equation holds for $\epsilon_0^c$.
The condition $[\hat{\rho}_{b\mp},\delta^{(1)}]A=0$ leads to
\begin{equation}
(\bar{\epsilon}_0\Gamma^{\mu_-})_{A_+}=0=(\bar{\epsilon}_0^c\Gamma^{\mu_-})_{A_+},
(\bar{\epsilon}_0\Gamma^{\mu_+})_{A_-}=0=(\bar{\epsilon}_0^c\Gamma^{\mu_+})_{A_-}.
\label{a}
\end{equation}
Similarly $[\hat{\rho}_{b\mp},\delta^{(1)}]B=0$, $[\hat{\rho}_{f\mp},\delta^{(1)}]\psi_i=0$
 and $[\hat{\rho}_{f\mp},\delta^{(2)}]\psi_0=0$ respectively yield
\begin{gather}
(\bar{\epsilon}_0\bar{\Gamma}^{(i_-)})_{A_+}=0
=(\bar{\epsilon}_0\bar{\Gamma}^{(i_+)})_{A_-},
\label{b}
\\
\begin{cases}
(\Gamma^{(j_-)}\Gamma^{(k_+)}\epsilon_0)_{A_-}
=(\Gamma^{\mu\pm}\bar{\Gamma}^{(i\mp)}\epsilon_0^c)_{A_-}=0,\\
(\Gamma^{(j_-)}\Gamma^{(k_-)}\epsilon_0)_{A_+}=
(\Gamma^{(j_+)}\Gamma^{(k_+)}\epsilon_0)_{A_+}=
(\Gamma^{\mu\mp}\bar{\Gamma}^{(i\mp)}\epsilon_0^c)_{A_+}=0,
\end{cases}
\label{pi}
\\
 {\rm and} \;\;\;   (\xi_0)_{A_-}=0.
\label{p02}
\end{gather}
Eqs.(\ref{p01})-(\ref{p02}) define a set of conditions satisfied 
by the anticommuting parameters $\epsilon_0$, $\xi_0$.

Let us find solutions to these equations.
The spinor $\epsilon_0$ is $\psi_0$type and must be of the form
\begin{equation}
\epsilon_0=(a,0,a,0,ia,0,-a,0,0,0,0,0,0,0,0,0)^t,
\end{equation}
where
$a=(\alpha,\beta)^t$. Similarly 
\begin{equation}
\epsilon_0^c=(0,0,0,0,0,0,0,0,0,ib,0,ib,0,b,0,-ib)^t,
\end{equation}
where
$b=(\gamma,\delta)^t$.
The spinor and vector indices are grouped into nonintersecting sets
${\mathcal A}_+$, ${\mathcal A}_-$, ${\mathcal M}_+$, ${\mathcal M}_-$,
${\mathcal I}_+$ and ${\mathcal I}_-$ such that
\[\mathcal{A}=\mathcal{A}_+\cup\mathcal{A}_-
=\{1,2,5,6,9,10,13,14,19,20,23,24,27,28,31,32\} \]
\[\mathcal{M}=\mathcal{M}_+\cup\mathcal{M}_-=\{0,1,2,3\},\mspace{20mu}
\mathcal{I}=\mathcal{I}_+\cup\mathcal{I}_-=\{1,2,3\}.\]
Let us first classify the possibilities by the division of $\mathcal M$ 
into $\mathcal M_+$ and $\mathcal M_-$.
This is done by using eq.(\ref{a}) and by following the procedure 
given in \cite{IT1}. 
It turns out  that there are three distinct possibilities for the division:
\begin{description}
\item[poss. 1.] \ ($\alpha\neq\beta,\alpha,\beta\neq0$;$\gamma\neq\delta,\gamma,\delta\neq0$);
$\{0,1,2,3\},\emptyset$,
\item[poss. 2.] \ ($\alpha=\pm\beta\neq0$;$\gamma=\pm\delta\neq0$);
\{0,1\},\{2,3\},
\item[poss. 3.] \ ($\alpha\neq0,\beta=0$\ or\ $\alpha=0,\beta\neq0$;$\gamma\neq0,\delta=0$\ or\ 
$\gamma=0,\delta\neq0$);
\{0,3\},\{1,2\}, 
\end{description}
Let us see each possibility more closely.
\begin{itemize}
\item poss. 1: 
From eq.(\ref{p01}), we see
\begin{equation}
\mathcal{A}_-=\mathcal{A},
\mspace{10mu}
\mathcal{A}_+=\emptyset,
\end{equation}
while $(\bar{\epsilon_0}\Gamma^{\mu_+})_{A_-}=0$ in eq.(\ref{a}) implies
\begin{equation}
\mathcal{M}_-=\mathcal{M},
\mspace{10mu}
\mathcal{M}_+=\emptyset.
\end{equation}
From eq.(\ref{b}), we conclude
\begin{equation}
\mathcal{I}_-=\mathcal{I},
\mspace{10mu}
\mathcal{I}_+=\emptyset.
\end{equation}
Finally eq.(\ref{p02}) tells us that the kinematical supersymmetry
 is broken completely:
\beq
\xi_0=0.
\eeq
This case has $4+0$ supersymmetries.

\item poss. 2: 
Following the same procedure as that of poss. 1, we conclude that this
 possibility does not lead to a consistent solution.
 
\item poss. 3:
This possibility leads to four different solutions. 

\noindent i) Choosing $a=(\alpha,0)^t,b=(\gamma,0)^t$, 
from eq.(\ref{p01}), we conclude
\begin{equation}
\mathcal{A}_-=\{1,5,9,13,19,23,27,31\},
\mspace{10mu}
\mathcal{A}_+=\{2,6,10,14,20,24,28,32\},
\end{equation}
while  from eq.(\ref{a}) and from eq.(\ref{b}),
 we conclude respectively
\beqn
\mathcal{M}_-=\{0,3\}, & &
\mathcal{M}_+=\{1,2\},  \;\;\\
  {\rm and} \;\;
\mathcal{I}_-=\{1,2,3\},  & &
\mathcal{I}_+=\emptyset.
\eeqn
Finally eq.(\ref{p02}) is solved by
\[\xi_0=(c,0,c,0,ic,0,-c,0,0,0,0,0,0,0,0,0)^t,\]
\[\xi_0^c=(0,0,0,0,0,0,0,0,0,id,0,id,0,d,0,-id)^t,\]
where $c=(0,\beta)^t,d=(0,\delta)^t$.

\noindent ii)Choosing $a=(0,\alpha)^t,b=(0,\gamma)^t$, 
from eq.(\ref{p01}), we conclude
\begin{equation}
\mathcal{A}_-=\{2,6,10,14,20,24,28,32\},
\mspace{10mu}
\mathcal{A}_+=\{1,5,9,13,19,23,27,31\},
\end{equation}
while from eq.(\ref{a}) and from eq.(\ref{b}),
 we respectively conclude
\beqn
\mathcal{M}_-=\{0,3\}, & &
\mathcal{M}_+=\{1,2\},  \;\;\\
  {\rm and} \;\;
\mathcal{I}_-=\{1,2,3\},  & &
\mathcal{I}_+=\emptyset.
\eeqn
Finally eq.(\ref{p02}) is solved by 
choosing $c=(\beta,0)^t,d=(\delta,0)^t$.

\noindent iii)Choosing $a=(\alpha,0)^t,b=(0,\gamma)^t$, 
from eq.(\ref{p01}), we conclude
\begin{equation}
\mathcal{A}_-=\{1,5,9,13,,20,24,28,32\},
\mspace{10mu}
\mathcal{A}_+=\{2,6,10,14,19,23,27,31\},
\end{equation}
while from eq.(\ref{a}) and from eq.(\ref{b}), 
we respectively conclude
\beqn
\mathcal{M}_-=\{0,3\}, & &
\mathcal{M}_+=\{1,2\},  \;\;\\
  {\rm and} \;\;
\mathcal{I}_-=\emptyset,  & &
\mathcal{I}_+=\{1,2,3\}.
\eeqn
Finally eq.(\ref{p02}) is solved by 
choosing $c=(0,\beta)^t,d=(\delta,0)^t$.

\noindent iv)Choosing $a=(0,\alpha)^t,b=(\gamma,0)^t$, 
from eq.(\ref{p01}), we conclude
\begin{equation}
\mathcal{A}_-=\{2,6,10,14,19,23,27,31\},
\mspace{10mu}
\mathcal{A}_+=\{1,5,9,13,20,24,28,32\},
\end{equation}
while from eq.(\ref{a}) and from eq.(\ref{b}), 
we respectively conclude
\beqn
\mathcal{M}_-=\{0,3\}, & &
\mathcal{M}_+=\{1,2\},  \;\;\\
  {\rm and} \;\;
\mathcal{I}_-=\emptyset,  & &
\mathcal{I}_+=\{1,2,3\}.
\eeqn
Finally eq.(\ref{p02}) is solved by
 choosing $c=(\beta,0)^t,d=(0,\delta)^t$.

\end{itemize}

These four cases have $2+2$ supersymmetries.

\section{Enumerating the Cases with Four or Eight Supercharges}
\label{5}
Let us generalize the results obtained in the last section.
We would first need to rewrite supersymmetry transformations 
in the new variables $A_{\mu}$, $B_i$, $\psi_0$ and $\psi_i$, but
we will not spell out its explicit form here.
As we have seen in the last section, 
the condition
$[\hat{\rho}_{b\mp}^{(\mu)},\delta^{(1)}]A_{\mu}=0$
yields 
\begin{align}
(\bar{\epsilon}_0\Gamma^{\mu_-})_{A_+}=
(\bar{\epsilon}_i\Gamma^{\mu_-})_{A_+}=
(\bar{\epsilon}_0^c\Gamma^{\mu_-})_{A_+}=
(\bar{\epsilon}_i^c\Gamma^{\mu_-})_{A_+}=0,\notag\\
(\bar{\epsilon}_0\Gamma^{\mu_-})_{A_-}=
(\bar{\epsilon}_i\Gamma^{\nu_+})_{A_-}=
(\bar{\epsilon}_0^c\Gamma^{\mu_+})_{A_-}=
(\bar{\epsilon}_i^c\Gamma^{\mu_+})_{A_-}=0,
\end{align}
and 
the condition  $[\hat{\rho}_{b\mp}^{(i)},\delta^{(1)}]B_i=0$
leads to
\begin{gather}
(\bar{\epsilon}_0\bar{\Gamma}^{(i_-)})_{A_+}=
(\bar{\epsilon}_{i_-}\bar{\Gamma}^{(i_-)})_{A_+}=
(\bar{\epsilon}_{j}^c\bar{\Gamma}^{(i_-)})_{A_+}=0,\notag\\
(\bar{\epsilon}_{0}\bar{\Gamma}^{(i_+)})_{A_-}=
(\bar{\epsilon}_{i_+}\bar{\Gamma}^{(i_+)})_{A_-}=
(\bar{\epsilon}_{j}^c\bar{\Gamma}^{(i_+)})_{A_-}=0,
\end{gather}
where $j\neq i$. Here the repeated indices are not to be summed over 
 unless stated  explicitly.
Similarly, from $[\hat{\rho}_{b\mp}^{(i)},\delta^{(1)}]B_i^{\dag}=0$
 we obtain
\begin{gather}
(\bar{\epsilon}_0^c\Gamma^{(i_-)})_{A_+}=
(\bar{\epsilon}_{i_-}^c\Gamma^{(i_-)})_{A_+}=
(\bar{\epsilon}_{j}\Gamma^{(i_-)})_{A_+}=0,\notag\\
(\bar{\epsilon}_{0}^c\Gamma^{(i_+)})_{A_-}=
(\bar{\epsilon}_{i_+}^c\Gamma^{(i_+)})_{A_-}=
(\bar{\epsilon}_{j}\Gamma^{(i_+)})_{A_-}=0.
\end{gather}
The condition $[\hat{\rho}_{f\mp}^{(0)(A)},\delta^{(1)}](\psi_0)_A=0$ yields
\begin{gather}
(\Gamma^{\mu_-\nu_+}\epsilon_0)_{A_-}=0,\notag\\
(\Gamma^{\mu_-\nu_-}\epsilon_0)_{A_+}=
(\Gamma^{\mu_+\nu_+}\epsilon_0)_{A_+}=0,
\end{gather}
\begin{gather}
|\epsilon_{i_-j_+k}|(\bar{\Gamma}^{(i_-)}\bar{\Gamma}^{(j_+)}\epsilon_k)_{A_-}=0,\notag\\
|\epsilon_{i_-j_-k}|(\bar{\Gamma}^{(i_-)}\bar{\Gamma}^{(j_-)}\epsilon_k)_{A_+}=
|\epsilon_{i_+j_+k}|(\bar{\Gamma}^{(i_+)}\bar{\Gamma}^{(j_+)}\epsilon_k)_{A_+}=0,
\end{gather}
\begin{gather}
(\Gamma^{\mu_-}\bar{\Gamma}^{(i_+)}\epsilon_{i_+}^c)_{A_-}=
(\Gamma^{\mu_+}\bar{\Gamma}^{(i_-)}\epsilon_{i_-}^c)_{A_-}=0,\notag\\
(\Gamma^{\mu_-}\bar{\Gamma}^{(i_-)}\epsilon_{i_-}^c)_{A_+}=
(\Gamma^{\mu_+}\bar{\Gamma}^{(i_+)}\epsilon_{i_+}^c)_{A_+}=0,
\end{gather}
\begin{equation}
(\epsilon_0)_{A_+}=0,
\end{equation}
while $[\hat{\rho}_{f\mp}^{(0)(A)},\delta^{(1)}](\psi_0^c)_A=0$
\begin{gather}
(\Gamma^{\mu_-\nu_+}\epsilon_0^c)_{A_-}=0,\notag\\
(\Gamma^{\mu_-\nu_-}\epsilon_0^c)_{A_+}=
(\Gamma^{\mu_+\nu_+}\epsilon_0^c)_{A_+}=0,\\
|\epsilon_{i_-j_+k}|(\Gamma^{(i_-)}\Gamma^{(j_+)}\epsilon_k^c)_{A_-}=0,\notag\\
|\epsilon_{i_-j_-k}|(\Gamma^{(i_-)}\Gamma^{(j_-)}\epsilon_k^c)_{A_+}=
|\epsilon_{i_+j_+k}|(\Gamma^{(i_+)}\Gamma^{(j_+)}\epsilon_k^c)_{A_+}=0,\\
(\Gamma^{\mu_-}\Gamma^{(i_+)}\epsilon_{i_+})_{A_-}=
(\Gamma^{\mu_+}\Gamma^{(i_-)}\epsilon_{i_-})_{A_-}=0,\notag\\
(\Gamma^{\mu_-}\Gamma^{(i_-)}\epsilon_{i_-})_{A_+}=
(\Gamma^{\mu_+}\Gamma^{(i_+)}\epsilon_{i_+})_{A_+}=0,\\
(\epsilon_0^c)_{A_+}=0.
\end{gather}
The condition
$[\hat{\rho}_{f\mp}^{(i)(A)},\delta^{(1)}](\psi_i)_A=0$ leads to
\begin{gather}
(\Gamma^{\mu_-\nu_+}\epsilon_i)_{A_-}=0,\notag\\
(\Gamma^{\mu_-\nu_-}\epsilon_i)_{A_+}=
(\Gamma^{\mu_+\nu_+}\epsilon_i)_{A_+}=0,
\end{gather}
\begin{gather}
|\epsilon_{ij_+k}|(\Gamma^{\mu_-}\Gamma^{(j_+)}\epsilon_k^c)_{A_-}=
|\epsilon_{ij_-k}|(\Gamma^{\mu_+}\Gamma^{(j_-)}\epsilon_k^c)_{A_-}=0,\notag\\
|\epsilon_{ij_-k}|(\Gamma^{\mu_-}\Gamma^{(j_-)}\epsilon_k^c)_{A_+}=
|\epsilon_{ij_+k}|(\Gamma^{\mu_+}\Gamma^{(j_+)}\epsilon_k^c)_{A_+}=0,
\end{gather}
\begin{gather}
|\epsilon_{ij_+k_-}|(\Gamma^{(j_+)}\Gamma^{(k_-)}\epsilon_0)_{A_-}=0,\notag\\
|\epsilon_{ij_+k_+}|(\Gamma^{(j_+)}\Gamma^{(k_+)}\epsilon_0)_{A_+}=
|\epsilon_{ij_-k_-}|(\Gamma^{(j_-)}\Gamma^{(k_-)}\epsilon_0)_{A_+}=0,
\end{gather}
\begin{gather}
(\Gamma^{\mu_-}\bar{\Gamma}^{(i_+)}\epsilon_0^c)_{A_-}=
(\Gamma^{\mu_+}\bar{\Gamma}^{(i_-)}\epsilon_0^c)_{A_-}=0,\notag\\
(\Gamma^{\mu_-}\bar{\Gamma}^{(i_-)}\epsilon_0^c)_{A_+}=
(\Gamma^{\mu_+}\bar{\Gamma}^{(i_+)}\epsilon_0^c)_{A_+}=0,
\end{gather}
\begin{gather}
|\epsilon_{i_-j_+k}|(\Gamma^{(i_-)}\bar{\Gamma}^{(j_+)}\epsilon_{i_-})_{A_-}=
|\epsilon_{i_+j_-k}|(\Gamma^{(i_+)}\bar{\Gamma}^{(j_-)}\epsilon_{i_+})_{A_-}=0,\notag\\
|\epsilon_{i_-j_-k}|(\Gamma^{(i_-)}\bar{\Gamma}^{(j_-)}\epsilon_{i_-})_{A_+}=
|\epsilon_{i_+j_+k}|(\Gamma^{(i_+)}\bar{\Gamma}^{(j_+)}\epsilon_{i_+})_{A_+}=0,
\end{gather}
\begin{equation}
(\epsilon_i)_{A_+}=0.
\end{equation}
The condition 
 $[\hat{\rho}_{f\mp}^{(i)(A)},\delta^{(1)}](\psi_i^c)_A=0$ leads to
\begin{gather}
(\Gamma^{\mu_-\nu_+}\epsilon_i^c)_{A_-}=0,\notag\\
(\Gamma^{\mu_-\nu_-}\epsilon_i^c)_{A_+}=
(\Gamma^{\mu_+\nu_+}\epsilon_i^c)_{A_+}=0,
\end{gather}
\begin{gather}
|\epsilon_{ij_+k}|(\Gamma^{\mu_-}\bar{\Gamma}^{(j_+)}\epsilon_k)_{A_-}=
|\epsilon_{ij_-k}|(\Gamma^{\mu_+}\bar{\Gamma}^{(j_-)}\epsilon_k)_{A_-}=0,\notag\\
|\epsilon_{ij_-k}|(\Gamma^{\mu_-}\bar{\Gamma}^{(j_-)}\epsilon_k)_{A_+}=
|\epsilon_{ij_+k}|(\Gamma^{\mu_+}\bar{\Gamma}^{(j_+)}\epsilon_k)_{A_+}=0,
\end{gather}
\begin{gather}
|\epsilon_{ij_+k_-}|(\bar{\Gamma}^{(j_+)}\bar{\Gamma}^{(k_-)}\epsilon_0^c)_{A_-}=0,\notag\\
|\epsilon_{ij_+k_+}|(\bar{\Gamma}^{(j_+)}\bar{\Gamma}^{(k_+)}\epsilon_0^c)_{A_+}=
|\epsilon_{ij_-k_-}|(\bar{\Gamma}^{(j_-)}\bar{\Gamma}^{(k_-)}\epsilon_0^c)_{A_+}=0,
\end{gather}
\begin{gather}
(\Gamma^{\mu_-}\Gamma^{(i_+)}\epsilon_0)_{A_-}=
(\Gamma^{\mu_+}\Gamma^{(i_-)}\epsilon_0)_{A_-}=0,\notag\\
(\Gamma^{\mu_-}\Gamma^{(i_-)}\epsilon_0)_{A_+}=
(\Gamma^{\mu_+}\Gamma^{(i_+)}\epsilon_0)_{A_+}=0,
\end{gather}
\begin{gather}
|\epsilon_{i_-j_+k}|(\bar{\Gamma}^{(i_-)}\Gamma^{(j_+)}\epsilon_{i_-}^c)_{A_-}=
|\epsilon_{i_+j_-k}|(\bar{\Gamma}^{(i_+)}\Gamma^{(j_-)}\epsilon_{i_+}^c)_{A_-}=0,\notag\\
|\epsilon_{i_-j_-k}|(\bar{\Gamma}^{(i_-)}\Gamma^{(j_-)}\epsilon_{i_-}^c)_{A_+}=
|\epsilon_{i_+j_+k}|(\bar{\Gamma}^{(i_+)}\Gamma^{(j_+)}\epsilon_{i_+}^c)_{A_+}=0,
\end{gather}
\begin{equation}
(\epsilon_i^c)_{A_+}=0.
\end{equation}
In addition, from $[\hat{\rho}_{f\mp}^{(A)},\delta^{(2)}](\psi_0)_A=0$ and 
$[\hat{\rho}_{f\mp}^{(A)},\delta^{(2)}](\psi_i)_A=0$, we obtain
\begin{equation}
(\xi_0)_{A_-}=(\xi_i)_{A_-}=0.
\end{equation}

Upon ${\bf Z}_3$ orbifolding, the number of surviving
supersymmetries is related to 
the number of $b_i$  such that $b_i=0$ is satisfied. 
The cases with $4+4$ supercharges have only one such $b_i$, and
we obtain the following four possibilities with $4+4$ supercharges:  
\begin{itemize} 
\item $b_0=0, b_1=-a_1, b_2=-a_2, b_3=-a_3\mspace{10mu}(a_1+a_2+a_3=0)$
\item $b_0=-a_1, b_1=0, b_2=a_3, b_3=a_2\mspace{10mu}(a_1-a_2-a_3=0)$
\item $b_0=-a_2, b_1=a_3, b_2=0, b_3=a_1\mspace{10mu}(a_1-a_2+a_3=0)$
\item $b_0=-a_3, b_1=a_2, b_2=a_1, b_3=0\mspace{10mu}(a_1+a_2-a_3=0)$.
\end{itemize}
The first one is the model which we already treated in the last
 section.

Similarly we construct the models with $8+8$ supercharges,
 which have two of vanishing  $b_i$.
There are six possibilities: 
\begin{itemize}
\item $b_0=b_1=0,a_1=0,b_2=-b_3=-a_2=a_3$
\item $b_0=b_2=0,a_2=0,b_1=-b_3=-a_1=a_3$
\item $b_0=b_3=0,a_3=0,b_1=-b_2=a_2=-a_1$
\item $b_1=b_2=0,a_3=0,b_0=-b_3=-a_1=-a_2$
\item $b_1=b_3=0,a_2=0,b_0=-b_2=-a_1=-a_3$
\item $b_2=b_3=0,a_1=0,b_0=-b_1=-a_2=-a_3$.
\end{itemize}
We collect these possibilities in the table: 
\[
\begin{array}{|c||c|c|c|c|} \hline
\text{supersymmetry} &  b_0 & b_1 & b_2 & b_3 \\\hline
\multicolumn{1}{|c||}{4+4}  & 0 & -a_1 & -a_2 & -a_3 \\\cline{2-5}
\multicolumn{1}{|c||}{} &  -a_1 & 0 & a_3 & a_2 \\\cline{2-5}
\multicolumn{1}{|c||}{} &  -a_2 & a_3 & 0 & a_1 \\\cline{2-5}
\multicolumn{1}{|c||}{} &  -a_3 & a_2 & a_1 & 0 \\\hline
\multicolumn{1}{|c||}{8+8}  & 0 & 0 & -a_2 & a_2 \\\cline{2-5}
\multicolumn{1}{|c||}{} &  0 & -a_1 & 0 & a_1 \\\cline{2-5}
\multicolumn{1}{|c||}{} &  0 & -a_1 & a_1 & 0 \\\cline{2-5}
\multicolumn{1}{|c||}{} &  -a_1 & 0 & 0 & a_1 \\\cline{2-5}
\multicolumn{1}{|c||}{} &  -a_1 & 0 & a_1 & 0 \\\cline{2-5}
\multicolumn{1}{|c||}{} &  -a_2 & a_2 & 0 & 0 \\\hline
\end{array}
\]
In each possibility, we need only to keep $\epsilon_i$ such that
  $b_i=0$ is satisfied. 
Note that the individual forms of $\epsilon_i$ are written as 
\begin{align}
\epsilon_0&=(a,0,a,0,ia,0,-a,0,0,0,0,0,0,0,0,0)^t,\label{e0}\\
\epsilon_1&=(b,0,b,0,-ib,0,-b,0,0,0,0,0,0,0,0,0)^t,\label{e1}\\
\epsilon_2&=(c,0,-c,0,ic,0,c,0,0,0,0,0,0,0,0,0)^t,\label{e2}\\
\epsilon_3&=(d,0,d,0,-id,0,d,0,0,0,0,0,0,0,0,0)^t,\label{e3}\\
\epsilon_0^c&=(0,0,0,0,0,0,0,0,0,e,0,e,0,-ie,0,-e)^t,\label{e0c}\\
\epsilon_1^c&=(0,0,0,0,0,0,0,0,0,f,0,-f,0,if,0,-f)^t,\label{e1c}\\
\epsilon_2^c&=(0,0,0,0,0,0,0,0,0,g,0,-g,0,-ig,0,g)^t,\label{e2c}\\
\epsilon_3^c&=(0,0,0,0,0,0,0,0,0,h,0,h,0,ih,0,h)^t,\label{e3c}
\end{align}
where $a,b,c,d,e,f,g,h$ are two component real vectors.
Consequently we consider the conditions on these remaining parameters.

It is noted that eqs.(\ref{e1})-(\ref{e3}) become proportional to eq.(\ref{e0})
once we flip signs in one or two entries.
 The same is true for eq.(\ref{e1c})-(\ref{e3c}), 
which become proportional to eq.(\ref{e0c}) with one or two sign flips.
This means that
 the calculation in the last section is also applicable to the remaining
 possibilities. From each of the four possibilities
 of ${\bf Z}_{3}$ orbifolding  with $4+4$ supercharges, 
we obtain one case with $4+0$ supercharges and
 four cases with $2+2$ supercharges
  upon matrix orientifolding.  There are in total twenty such cases.

Likewise, to each of the six possibilities of ${\bf Z}_{3}$ orbifolding with $8+8$
 supercharges, we first find 
 an appropriate intersection of  above eqs.(\ref{e0})-(\ref{e3})
(\ref{e0c})-(\ref{e3c}) and impose the conditions of matrix orientifolding.
  In this way, we are able to exhaust all
 cases with
  either $8+0$ supersymmetries or $4+4$ supersymmetries
 upon ${\bf Z}_{3}$ orbifolding followed by matrix orientifolding.
  To each of the six possibilities, there exist one case with
 $8+0$ supersymmetries and four cases with  $4+4$ supersymmetries.
There are thirty such cases in total.
 
We conclude that there are in total fifty cases carrying
  four or eight supercharges upon ${\bf Z}_3$ orbifolding followed
 by  matrix orientifolding. 
\label{5}

\newpage
\section*{Acknowledgements}
This work is supported in part by the Grant-in-Aid for Scientific
Research(16540262) from the Ministry of Education,
Science and Culture, Japan.
Support from the 21 century COE program
``Constitution of wide-angle mathematical basis focused on knots"
is gratefully appreciated.

\appendix

\newpage

\end{document}